\documentclass[fleqn,usenatbib]{mnras}

\usepackage{newtxtext,newtxmath}

\usepackage[T1]{fontenc}

\DeclareRobustCommand{\VAN}[3]{#2}
\let\VANthebibliography\thebibliography
\def\thebibliography{\DeclareRobustCommand{\VAN}[3]{##3}\VANthebibliography}

\usepackage{graphicx}	
\usepackage{amsmath}	

\newcommand{\oiii}{\mbox{[O\,{\sc iii}]}}

\usepackage{ulem}

\title[FOREVER22: Star formation and clustering properties of protoclusters]{FOREVER22: Insights into star formation and clustering properties of protoclusters from simulations and JWST}

\author[K. Morokuma-Matsui et al.]{
Kana Morokuma-Matsui,$^{1,2}$\thanks{E-mail: kanamoro@ioa.s.u-tokyo.ac.jp}
Hidenobu Yajima,$^{2}$
and Makito Abe$^{2,3}$
\\
$^{1}$Institute of Astronomy, Graduate School of Science, The University of Tokyo, 2-21-1 Osawa, Mitaka, Tokyo 181-0015, Japan\\
$^{2}$Center for Computational Sciences, University of Tsukuba, Ten-nodai, 1-1-1 Tsukuba, Ibaraki 305-8577, Japan\\
$^{3}$Faculty of Natural Sciences, National Institute of Technology, Kure College, 2-2-11 Agaminami, Kure, Hiroshima 737-8506, Japan\\
}

\date{Accepted XXX. Received YYY; in original form ZZZ}

\pubyear{2015}

\begin{document}
\label{firstpage}
\pagerange{\pageref{firstpage}--\pageref{lastpage}}
\maketitle

\begin{abstract}
Using cosmological hydrodynamic simulations with radiative transfer, we investigate star formation and overdensity ($\delta$) in Coma-type cluster progenitors from $z=14$ to 6. Our simulations reproduce observed $M_{\rm star}$-SFR relations and $\delta$ at these redshifts. We find: (1) protocluster (PC) and mean-density field (MF) galaxies show similar $M_{\rm star}$-SFR relations, with PC galaxies extending to higher $M_{\rm star}$ and SFR. (2) UV-bright PC galaxies ($M_{\rm UV}\lesssim -20$~mag) have $>2$ mag higher UV attenuation and shallower UV slopes than MF galaxies. (3) $\delta$ increases with redshift, depending on observational parameters (e.g., $\delta\sim50$ at $z=14$ to $\delta\sim3$ at $z=6$ for a search volume of $\sim3000$~cMpc$^3$ and a limiting magnitude of $M_{\rm UV}=-17$~mag). These results indicate that enhanced star formation in PCs is driven by massive galaxy overdensity, not anomalously high specific SFR. While simulated $\delta$ agrees with observed PC candidates (potential Coma progenitors), some MF galaxies show comparable $\delta$. We propose a robust PC identification method using both $\delta$ and $M_{\rm star}$ of the most massive member. Critical $M_{\rm star}$ thresholds for Coma progenitors are estimated ($10^{7.1}$ to $10^{10.2}$ M$_\odot$ from $z=14$ to 6). Comparison with JWST observations suggests GS-z14-0 and GS-z14-1, the current highest redshift holders, are likely progenitors of Coma-type clusters.

\end{abstract}




\section{Introduction}

Protoclusters (PCs) are overdense regions in the early Universe destined to become today's galaxy clusters.
As progenitors of massive structures, PCs are believed to have initiated cosmic reionization \citep[e.g.,][]{McQuinn:2007rc,Weinberger:2018rp,Yajima:2022jl} and driven the chemical evolution of the Universe \citep[e.g.,][]{Fukushima:2022ky}.
Theoretical models predict that PCs contribute significantly to the cosmic star formation rate density (CSFRD), with their fraction increasing to 20-50\% at redshifts around five according to recent theoretical models \citep{Chiang:2017xz, Lim:2024um}, which is supported by recent observations \citep{Sun:2024pb}.
Understanding the star formation properties within these massive PCs is crucial for elucidating the star formation processes in the early Universe.

Deep, ground-based spectroscopic surveys have identified PC candidates at redshifts of $z\sim6-7$ using Lyman-$\alpha$ emission \citep[e.g.,][]{Toshikawa:2012tm,Toshikawa:2014zm,Chanchaiworawit:2017fx,Chanchaiworawit:2019om,Harikane:2019ss}.
Recent observations with the James Webb Space Telescope (JWST) have significantly expanded the sample of spectroscopically confirmed PC candidates beyond redshift $z>6$ \citep[e.g.,][]{Laporte:2022rf,Morishita:2023jh,Helton:2024fx,Arribas:2024dt,Wang:2023ph,Fudamoto:2025hx}.
\cite{Laporte:2022rf} and \cite{Morishita:2023jh} discovered PC candidates at $z\sim8$ behind the galaxy clusters, SMACS~0723-7327 (``SMACS0723\_PC'') and Abell~2744 (``A2744-z7p9OD''), respectively.
\cite{Hashimoto:2023tg} identified four star-forming galaxies within a compact region of $11$~kpc~$\times$~$11$~kpc at the core of A2744-z7p9OD, suggesting the potential formation of a massive galaxy within 100~Myr.
This discovery hints at the early stages of forming the brightest cluster galaxies observed in the present universe.
\cite{Helton:2024fx} identified $12$ PC candidates at $z>6$ based on H$\alpha$ and \oiii~emission.
They reported a PC number density of $n_{\rm PC}\sim2.2\times10^{-5}$~comoving~Mpc$^{-3}$ (cMpc$^{-3}$) at the Epoch of Reionization (EoR), significantly exceeding predictions for Coma-type and Fornax-type clusters based on semi-analytic galaxy formation models \citep{Chiang:2013eb}.
This indicates that either the observational data overestimates the number of protoclusters or the simulations underestimate their formation.

Observational studies of PCs are hindered by a number of significant challenges as outlined in \cite{Lim:2024um}. While PCs are theoretically defined well by cosmological simulations following the time evolution of large-scale structure, their observational identification is complex due to the reliance on overdense regions of various galaxy populations, including Lyman-$\alpha$ emitters (LAEs), Lyman-break galaxies (LBGs), H$\alpha$ emitters (HAEs), dusty star-forming galaxies (DSFGs), sub-millimeter galaxies (SMGs), quasi-stellar objects (QSOs), high-redshift radio galaxies (HzRGs), and Lyman-$\alpha$ blobs (LABs).
The choice of tracer galaxies introduces substantial variability in sample selection and analysis. Moreover, defining PC boundaries and estimating their properties is hampered by observational limitations and the inherent complexity of these systems.
\cite{Lim:2024um} further emphasizes the significant impact of aperture choice on total mass estimates within PCs, with uncertainties reaching an order of magnitude.

To reveal the relationship between the natures of PCs and observable properties, we investigate the formation of galaxies and their spatial distributions of PCs modeled in cosmological simulations.
We utilized the large-scale cosmological hydrodynamic simulation FOREVER22 \citep{Yajima:2022jl} to examine galaxy properties.
This simulation features a $(714.2~{\rm cMpc})^{3}$ N-body simulation volume and a baryonic mass resolution of $2.9\times10^6$~M$_\odot$ in zoom-in hydrodynamical simulations.
\cite{Yajima:2023tn} and \cite{Harada:2023qt} explored Population III stars and metal enrichment within the FOREVER22 framework.
To enable direct comparison with observations, we perform radiative transfer calculations using the {\tt ART$^2$} code, incorporating dust extinction effects \citep{Yajima:2012ty,Li:2020ez}.
Our analysis focuses on the following key aspects: the star formation main sequence (SFMS) of PC galaxies, the evolution of galaxy overdensity, and the properties of the UV continuum. By examining these properties, we aim to shed light on the formation and growth of PCs, the environmental impact on galaxy evolution, and the potential observational signatures of these early structures.

This paper is organized as follows. Section~\ref{sec:forever22art2} provides a brief overview of the simulations and radiative transfer calculations. Section~\ref{sec:resultsdiscussions} presents our results on the galaxy overdensity and star formation properties of protocluster galaxies. We also discuss potential methods for identifying protoclusters based on these findings in section~\ref{sec:dis_pc}. Finally, section~\ref{sec:summary} summarizes our conclusions.

\section{Cosmological simulation and radiative transfer}\label{sec:forever22art2}

\subsection{Cosmological hydrodynamic simulation-- FOREVER22}

\begin{figure*}
\begin{center}
\includegraphics[width=\textwidth, bb=0 0 1741 1195]{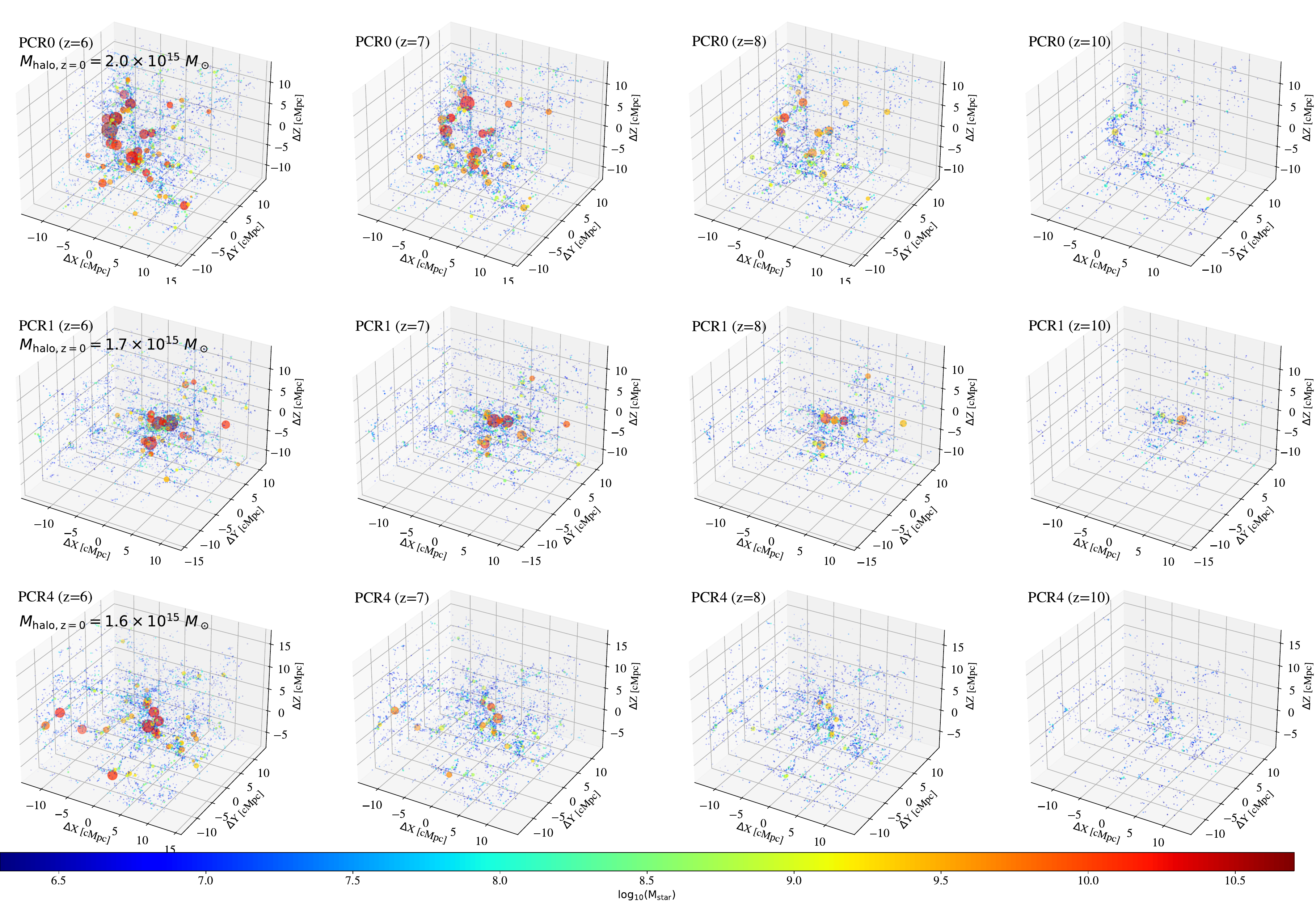}
\end{center}
\caption{
Three-dimension (3D) distribution of galaxies in the PCR0, 1, and 4 at $z=6,7,8,10$.
The coordinates represent distances from the barycenter within the zoom-in box.
The symbol size scales SFR and the color indicates stellar mass.
From top to bottom, the distributions of galaxies in PCR0, PCR1, and PCR4 are presented.
From left to right, the redshift increases from $z=6$ to $z=10$.
The final halo mass at $z=0$ for each PC is also presented in the left-hand side panel ($z=6$ panel).
}
\label{fig:3d}
\end{figure*}

\begin{figure}
\begin{center}
\includegraphics[width=0.5\textwidth, bb=0 0 509 707]{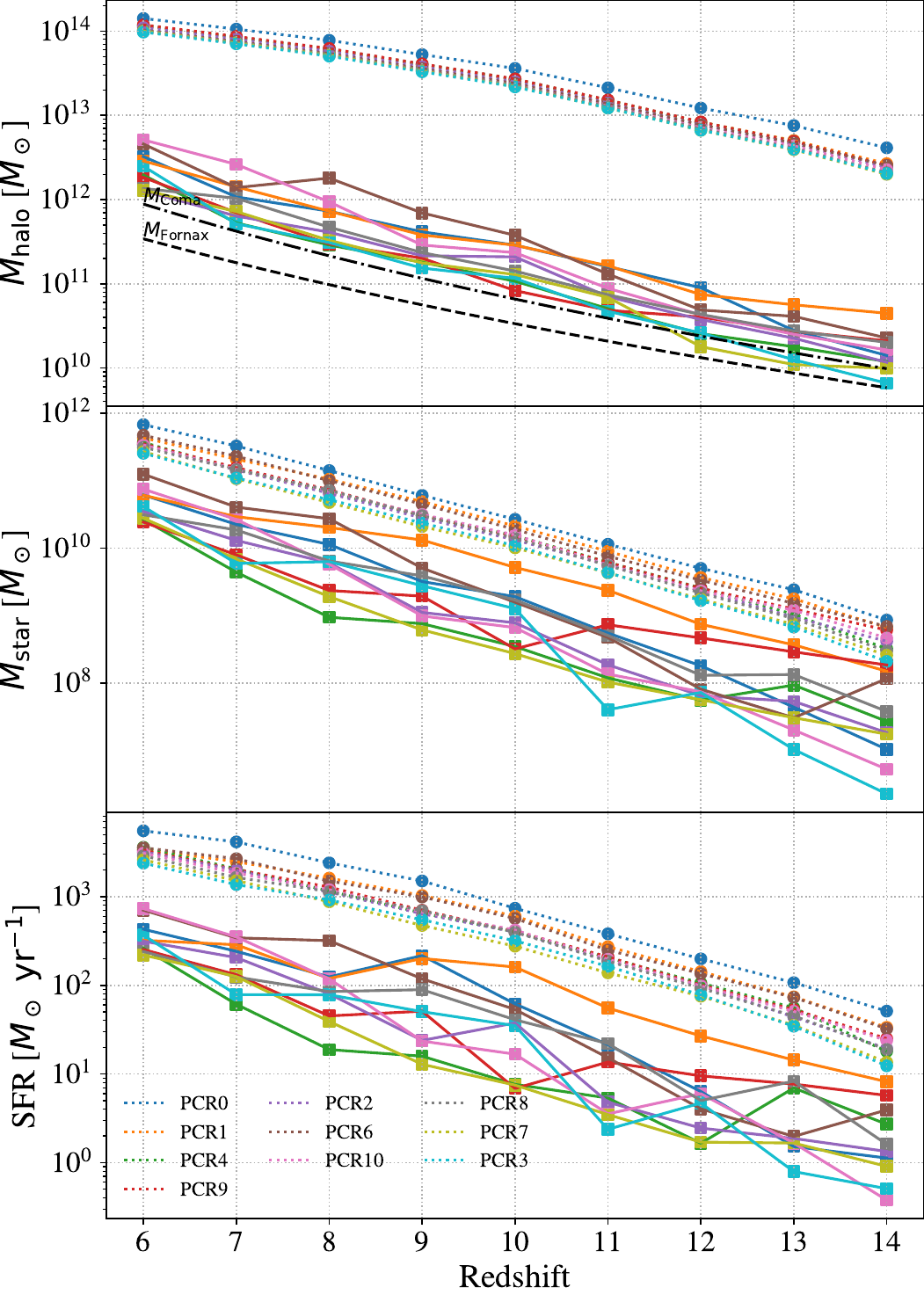}
\end{center}
\caption{
Redshift evolution of halo mass (top), stellar mass (middle), and SFR (bottom) of a BCG at each redshift and entire galaxies within the PCR calculation box.
The halo mass of all the PCRs reaches $10^{14}$~M$\odot$ at $z=6$.
}
\label{fig:z_evo}
\end{figure}

We here briefly explain the FOREVER22 project \citep{Yajima:2022jl}.
The FOREVER22 project employs the SPH code {\tt GADGET-3} \citep{Springel:2005kh}, modified for the OWLS \citep{Schaye:2010nw}, and further extended for the FiBY project \citep{Johnson:2013uk,Paardekooper:2015ft} to include Pop III star formation, Lyman–Werner feedback, and non-equilibrium primordial chemistry. Metal line cooling follows \cite{Wiersma:2009ng}.
We modified the star formation, black hole accretion, and supernova feedback models based on {\tt EAGLE} \cite{Schaye:2015lq} and implemented the radiative feedback from young stars, kinetic feedback from black holes, and dust grain evolution into the code.
The FOREVER22 project covers a wide range of cosmological scales using three different zoom set-ups in a parent volume of $(714.2~{\rm cMpc})^{3}$: PCR, BCG, and First.
While observations reveal both large-scale protocluster structures \citep{Kikuta:2019al} and small-scale galaxy details \citep{Tadaki:2018hn}, simulations struggle to capture both scales simultaneously. Our simulations aim to address this by investigating the statistical properties and detailed structure of protocluster galaxies.

For the PCR run analyzed in this paper, we employ $(28.6~\mathrm{cMpc})^3$ simulation volumes for each PCR to investigate the statistical properties of galaxies within protoclusters. The top 10 most massive halos are identified at $z=2$ from the parent N-body calculation whose volume is $(714.2~{\rm cMpc})^{3}$, for the zoom-in high-resolution calculations.
All simulations produce 200 snapshots from $z=100$ to the final redshift. While our zoom-in regions cover the typical volume for a galaxy cluster at $z=0$ \citep{Chiang:2017xz}, massive clusters require larger volumes \citep{Muldrew:2015ns,Lovell:2018lc}. We find that our PCRs form massive haloes ($>10^{14}$~$M_\odot$) and enclose $46-80$~\% of their descendant cluster's mass at $z=0$. Despite this limitation, we capture the main progenitors and massive galaxies crucial for our study of galaxy evolution in overdense environments.

For comparison, we also performed zoom-in simulations of mean-density fields (hereafter, MF).
We randomly selected three distinct $(28.6~\mathrm{cMpc})^3$ regions from the parent N-body simulation, ensuring they exhibited a normal halo mass function and matter density.
The MF zoom-in simulations used the same mass and spatial resolutions as the PCR runs.

The 3D distribution of galaxies in each PCR is shown in Fig.~\ref{fig:3d}.
There are hub-filament structures in all PCRs as seen around the PCs at $z=6-7$ \citep[][]{Harikane:2019ss}.
The number of galaxies with large stellar masses increases with time.
Massive galaxies, indicated by redder colors, predominantly occupy the cores of PCR1 and PCR2. However, such massive galaxies are also distributed within the filamentary structures of PCR0.

\subsection{Radiative transfer --{\tt ART$^2$} \label{sec:art2}}

Massive galaxies in PCs can be metal and dust enriched even at $z \gtrsim 6$ because the star formation proceeds earlier. Therefore, a part of stellar radiation can be absorbed by interstellar dust. To reasonable comparison with observation, we here perform radiative transfer calculations with {\tt ART$^2$} \citep{Li:2008me, Yajima:2012ty}.
{\tt ART$^2$} is a 3D Monte Carlo radiative transfer code that calculates continuum emission from X-ray to radio and utilizes an adaptive refinement grid structure. The code also includes modeling metal and CO lines, Ly$\alpha$ line, and the ionization of neutral hydrogen.

\section{Galaxy overdensity and star formation properties}
\label{sec:resultsdiscussions}

In the following subsections, we investigate galaxy properties within overdense regions and analyze the evolution of overdensity in the FOREVER22 simulations.
Specifically, we present:
The redshift evolution of progenitor halos of present-day Coma clusters (Section~\ref{sec:redshiftevolution});
The stellar mass-star formation rate (SFR) relation of galaxies in PCs (Section~\ref{sec:mstarsfr});
The dust attenuation properties of galaxies in PCs (Section~\ref{sec:muv});
The evolution of galaxy overdensity at redshifts $z\sim6-14$ (Section~\ref{sec:overdensity}).

\subsection{Redshift evolution of $M_{\rm halo}$, $M_{\rm star}$ and SFR of the Coma-cluster progenitors}\label{sec:redshiftevolution}

Fig.~\ref{fig:z_evo} illustrates the redshift evolution of integrated halo mass ($M_{\rm halo}$), stellar mass ($M_{\rm star}$), and SFR within each entire PCR simulation box (dotted lines) and within the most massive halo within each PCR (solid lines).
For comparison, we also plot the redshift evolution of $M_{\rm halo}$ for the progenitors of Coma- ($M_{\rm Coma}$) and Fornax-type galaxy clusters ($M_{\rm Fornax}$), calculated using the extended Press-Schechter formalism following \cite{Neistein:2008cr,Dekel:2013bh}.
PCs hosting galaxies with $M_{\rm halo}>M_{\rm Coma}$ or $M_{\rm halo}>M_{\rm Fornax}$ at a given redshift are likely to evolve into Coma- or Fornax-type galaxy clusters at $z=0$, respectively.
\cite{Yajima:2022jl} confirmed that PCs in the FOREVER22 simulations grow to become as massive as the Coma cluster by the present day ($z=0$).
We find that SPT0311-58, one of the extreme PC candidates at $z\sim6.9$,  hosts a massive core with a halo mass of $M_{\rm halo}\sim 5\times 10^{12}$~$M_\odot$ \citep{Arribas:2024dt}, comparable to or slightly exceeding values in our simulations.

Observational studies often incorrectly compare the integrated halo masses of galaxies within PCs to theoretical predictions for a single progenitor halo of present-day galaxy clusters, as highlighted by \cite{Lim:2024um}.
This comparison is problematic because PCs are dynamically evolving systems with ill-defined boundaries.
Furthermore, accurately estimating the total halo mass of PCs is challenging due to the significant contribution from low-mass galaxies, which often remain undetected by current observational surveys.
A more promising approach is to focus on the most massive member galaxy within the PC, and compare its mass to theoretical predictions. This approach provides a more robust and easier observational comparison by focusing on a well-defined and readily observable component of the PCs.

\subsection{$M_{\rm star}-$SFR relation}\label{sec:mstarsfr}

Galaxies shape so-called the ``main sequence'' of star-forming galaxies (hereafter, SFMS) on the $M_{\rm star}-$SFR plane, and the SFMS evolves with time \cite[e.g.,][]{Speagle:2014ow}.
Recent JWST observations of galaxies in PCs at the EoR find that most galaxies follow the SFMS at each redshift \citep[e.g.,][]{Laporte:2022rf,Morishita:2023jh,Helton:2024fx}.

\subsubsection{Comparison of PCR and MF galaxies}\label{sec:mstarsfr_pcrmf}

\begin{figure*}
\begin{center}
\includegraphics[width=\textwidth, bb=0 0 1417 1377]{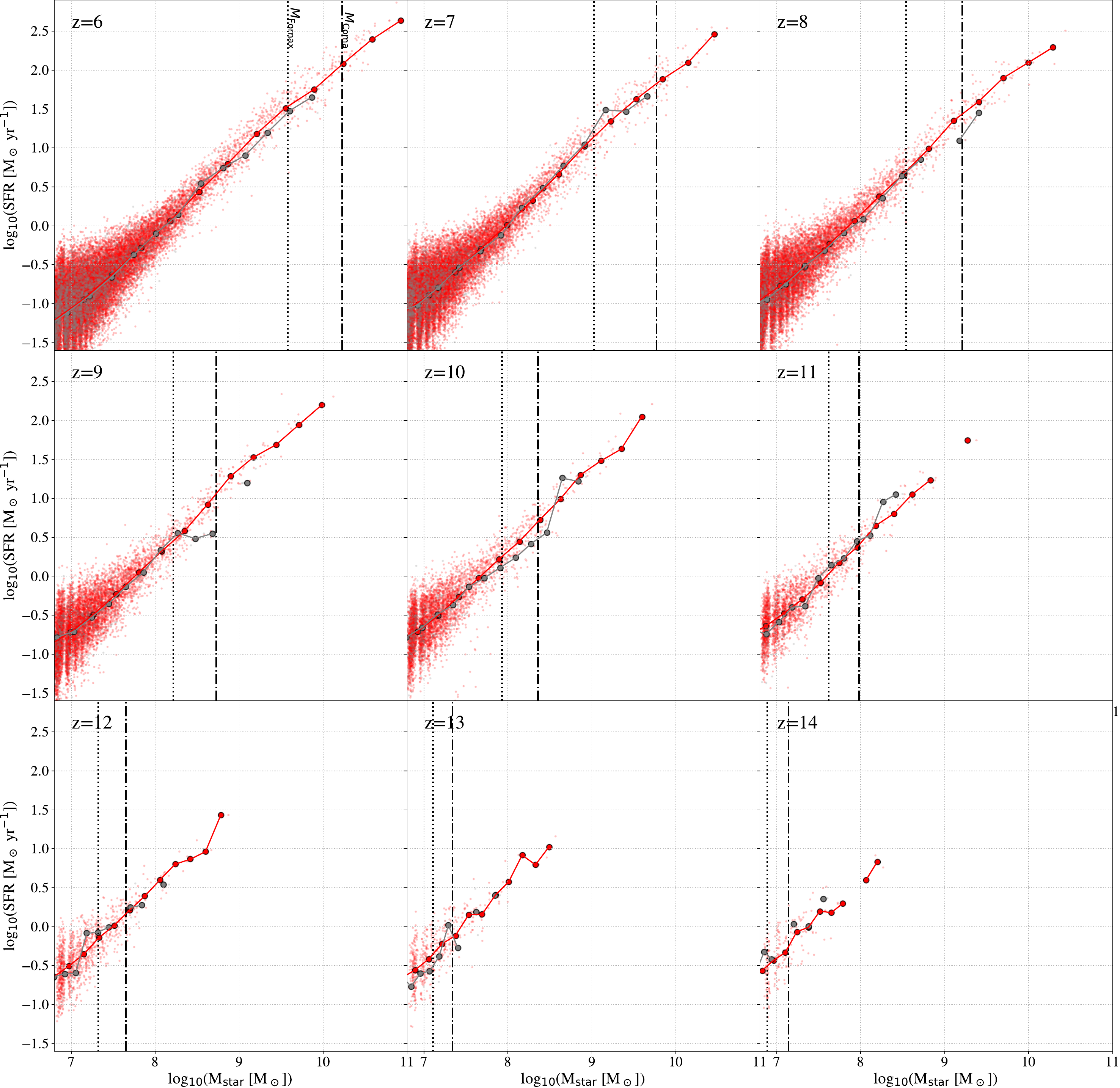}
\end{center}
\caption{
Comparison of the $M_{\rm star}-$SFR relation of the PC galaxies and MF galaxies at $z=6-14$.
Individual PCR and MF galaxies are indicated with open red and gray circles, respectively.
The filled red and gray circles indicate the medians of SFR for galaxies with similar $M_{\rm star}$.
Dotted and dot-dashed lines indicate the progenitor stellar masses of Fornax-type ($M_{\rm Fornax, \star}$) and Coma-type galaxy clusters ($M_{\rm Coma, \star}$) at each redshift, respectively.
}
\label{fig:mstarsfr_mf}
\end{figure*}

We find that the SFMSs of the PCR and the MF galaxies appear almost identical for all the redshifts analyzed here.
We compare the $M_{\rm star}-$SFR relation between the PCR and MF galaxies in Fig.~\ref{fig:mstarsfr_mf}.
The relation is more extended to the higher $M_{\rm star}$ and SFR regime for the PC galaxies compared to the MF galaxies.
Thus, the more active star formation in the PC galaxies compared to the MF galaxies is attributed to the high fraction of massive galaxies rather than the high specific SFR in PCs.
The high fraction of massive galaxies in PC, consequently high-SFR galaxy, has already been claimed in \cite{Yajima:2022jl}.
They found that, when comparing the stellar mass functions (SMF) of PCR and MF galaxies, PCR galaxies exhibit an excess at high stellar mass regimes and the normalization of the SMF of the PCR galaxies is higher than those of the MF galaxies by a factor of $\gtrsim 3$ at $z\gtrsim 2$.
They interpreted these differences as attributed to an efficient galaxy merger and an accelerated galaxy evolution in PCRs.

The high fraction of galaxies with high $M_{\rm star}$ and SFR in the PC region is also reported in the observational studies \citep[e.g.,][]{Larson:2022km,Leonova:2022qw,Tang:2023ml,Toshikawa:2024kt}.
\cite{Toshikawa:2024kt} photometrically identified 111 PC candidates at $3<z<5$ using the dropout method based on the data obtained in the Hyper SuprimeCam Subaru Strategic Programme and the CFHT Large Area U-band Deep Survey.
They found that UV-bright galaxies are overabundant in the PC region.

\subsubsection{Comparison between FOREVER22 and observations}\label{sec:mstarsfr_obs}

\begin{figure*}
\begin{center}
\includegraphics[width=\textwidth, bb=0 0 1417 1377]{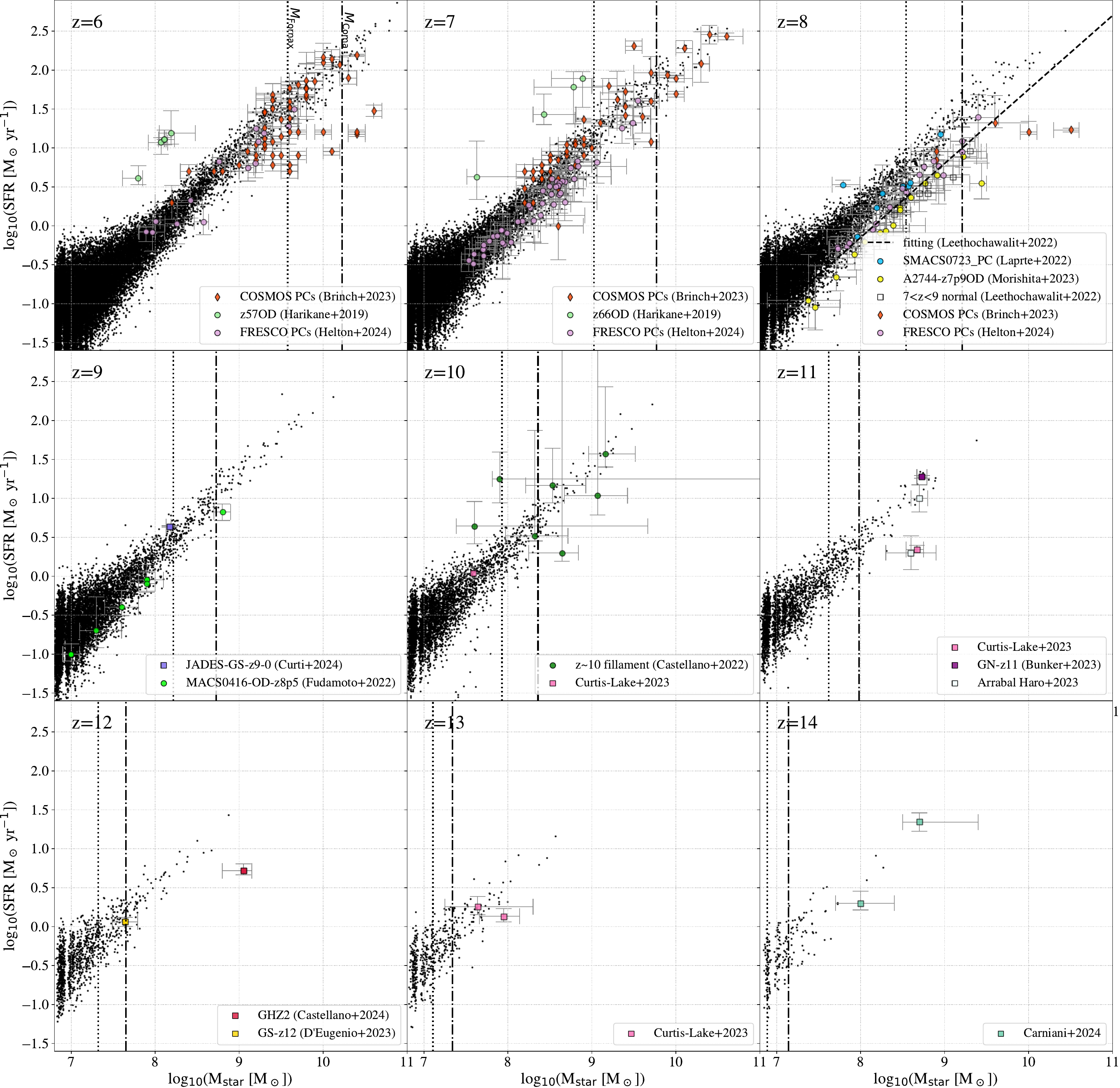}
\end{center}
\caption{
$M_{\rm star}-$SFR relation of galaxies in the PC regions.
Observation data of galaxies in spectroscopically identified overdense regions \citep{Harikane:2019ss,Laporte:2022rf,Morishita:2023jh,Castellano:2023ra,Helton:2024fx,Fudamoto:2025hx} are indicated with circles.
Galaxies in overdense regions identified based on the photo-$z$ information in the COSMOS field are indicated with diamonds at $z=6-8$ \citep{Brinch:2023mq}.
Higher-$z$ galaxies that have not been confirmed to be in the overdense regions are indicated with squares \citep{Leethochawalit:2023vz,Curti:2024xr,Bunker:2023cz,Arrabal-Haro:2023li,Castellano:2024zk,DEugenio:2023zp,Carniani:2024hb}.
Progenitor $M_{\rm star}$ for the brightest-cluster galaxies in the Coma-type and Fornax-type clusters are shown with dot-dash and dotted lines, respectively.}
\label{fig:mstarsfr}
\end{figure*}

We compare the $M_{\rm star}-$SFR relation from our simulations to several observations in Fig.~\ref{fig:mstarsfr}.
We mainly consider the spectroscopically identified PC candidates \citep{Laporte:2022rf,Morishita:2023jh,Helton:2024fx,Fudamoto:2025hx} while we also show the PC candidates identified with the photo-$z$ ($z_{\rm p}$) data sample in the COSMOS field \citep{Brinch:2023mq} and a filamentary structure at $z\sim10$ \citep{Castellano:2023ra}.
The error of the $z_{\rm p}$ estimates of galaxies in the COSMOS field is $\Delta z_{\rm p} \sim 0.1-0.2$, which corresponds to $43-85$~cMpc at $z\sim 6$, $35-70$~cMpc at $z\sim 7$, and $30-59$~cMpc at $z\sim 8$.
In the Castellano's filament, one of the member galaxies is spectroscopically confirmed.
We also plot spectropically-identified galaxies at $z>9$ without a confirmation of their overdense environments \citep{Curti:2024xr,Curtis-Lake:2023ol,Bunker:2023cz,Arrabal-Haro:2023li,Castellano:2024zk,DEugenio:2023zp,Carniani:2024hb}.

Although simulations often underestimate star-formation activity compared to observations \citep{Lim:2021vx}, our simulation demonstrates good overall agreement with observational data, potentially due to improved baryonic mass resolution.
However, discrepancies remain.
Specifically, observations reveal galaxies with suppressed SFRs for their $M_{\rm star}$.
In contrast, LAEs in PC regions, as reported by \cite{Harikane:2019ss}, exhibit systematically higher SFRs than the SFMS at $z=6$ and 7.
At $z=8$, our simulated galaxies show slightly elevated SFRs compared to observed galaxies of similar masses.
Notably, observations indicate the presence of galaxies with decreased SFRs at fixed Mstar even at $z=11-12$.

\subsubsection{Stellar mass of the most massive member galaxy, as a signpost candle of PC}\label{sec:mstarsfr_progenitor}

\begin{figure}
\begin{center}
\includegraphics[width=0.5\textwidth, bb=0 0 494 487]{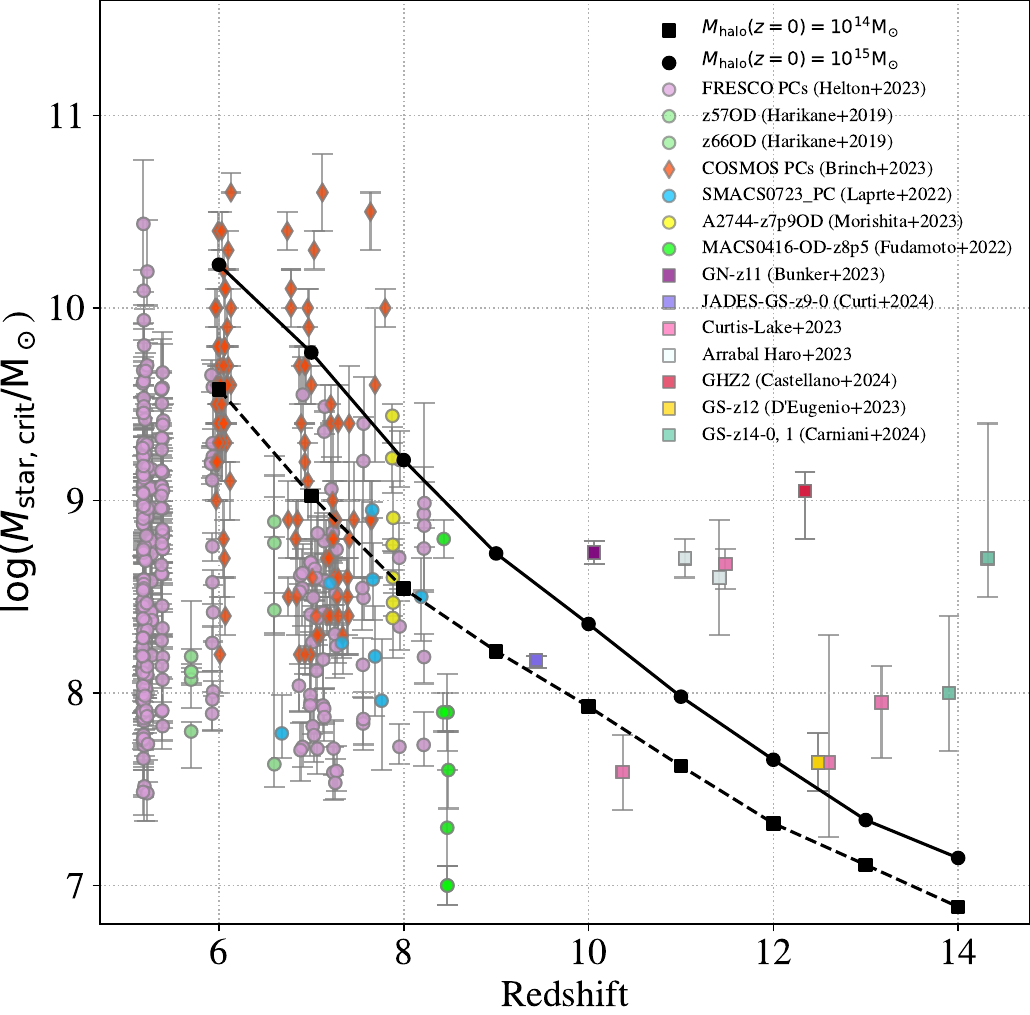}
\end{center}
\caption{
$M_{\rm star,crit}$ evolution for progenitors of Coma- and Fornax-class galaxy clusters.
$M_{\rm star,crit}$ is estimated based on the extended Press-Schechter formalism and the $M_{\rm halo}-M_{\rm star}$ relation derived with FOREVER22 data.
Observational data for the member galaxies within the overdense regions are also presented using the same symbols in Fig.~\ref{fig:mstarsfr}.
}
\label{fig:mstarcrit}
\end{figure}

We also show stellar masses ($M_{\rm star,crit}$) for the progenitors of Coma- and Fornax-type galaxy clusters in Fig.~\ref{fig:mstarsfr} and Fig.~\ref{fig:mstarcrit}, which are estimated with the $M_{\rm halo}$ evolution expected from the extended Press-Schechter formalism \citep{Neistein:2008cr,Dekel:2013bh} as shown in Fig.~\ref{fig:z_evo} and the $M_{\rm halo}-M_{\rm star}$ relationship from the FOREVER22 data (Fig.~\ref{fig:mhalomstar}).
Here we adopted $M_{\rm halo}$ of $10^{15}$~M$\odot$ and $10^{14}$~M$_\odot$ at $z=0$ for Coma- and Fornax-type galaxy clusters, respectively.
PCs hosting galaxies with $M_{\rm star}$ higher than these lines are more likely to evolve to Coma- or Fornax-type galaxy clusters than those with a lower $M_{\rm star}$.

We can see that some observed galaxies in PC candidates, especially found with the COSMOS photo-$z$ catalog, i.e., based on the LBGs methods, have $M_{\rm star}$ higher than $M_{\rm Coma}$, suggesting they are progenitors of the Coma-type clusters.
On the other hand, galaxies in the FRESCO PCs, z57OD, and z66OD generally have $M_{\rm star}\lesssim M_{\rm Coma}$ at $z=6-7$ while some FRESCO galaxies in the JADES-GS-OD-7.561 and JADES-GS-OD-7.954 have $M_{\rm star}>M_{\rm Coma}$ at $z=8$ \citep{Helton:2024fx}.
This suggests the majority of FRESCO PCs, z57OD, and z66OD may not be a progenitor of the Coma-type clusters.
It should be noted that these ovserdensities are claimed to have the total $M_{\rm halo}$ of galaxies within the PC region exceeding the expected $M_{\rm halo}$ for the Coma-type clusters, however, it is not clear whether these galaxies reside in a single halo.
It is still possible that the most massive galaxies are not found yet in these ovserdensities without galaxies with $M_{\rm star}>M_{\rm Coma}$ since these overdensities are identified with emission lines at UV or optical wavelengths.
It is interesting to search for possible member galaxies by spectroscopically detecting their Lyman breaks and by observing the dust continuum to identify massive member galaxies without active star formation and dusty star-forming galaxies, respectively.

On the other hand, almost all galaxies with the spectroscopic redshift measurements at $z>10$ have $M_{\rm star}>M_{\rm Coma}$, although they are not confirmed to be in the overdense regions yet.
As we and other previous studies have found, massive galaxies preferentially occur in overdense regions.
Thus, we might have just observed the tip of the iceberg of the PCs.
GN-z11 is an unusually luminous object at $z=10.6$ with $M_{\rm UV}\sim-21.6$~mag compared to the luminosity function at $z\sim10-11$ \citep[e.g.,][]{Perez-Gonzalez:2023ov}.
Its stellar mass is estimated to be $M_{\rm star}=10^{8.73}$~M$_\odot$ \citep{Bunker:2023cz}, which is higher than $M_{\rm Coma}$ at its redshift.
This may suggest that GN-z11 resides in the overdense region.
Based on the JWST/NIRCam observations, \cite{Tacchella:2023zp} found nine galaxies with photometric redshifts of $9.6-10.8$ within $\sim5$~cMpc around GN-z11.
Later, \cite{Scholtz:2023rp} calculated $\delta$ of $\gtrsim 27$ based on four spectroscopically confirmed galaxies around GN-z11.
Although the estimated total halo mass of the GN-z11 system of $2.96\times10^{10}$~M$_\odot$ in their study is smaller than those in our modeled PCs. 
there might be other member galaxies and they would evolve into clusters as massive as the Coma cluster.
GS-z14-0 at $z=14.18$ \citep{Carniani:2024hb, Carniani:2024vh}, the galaxy with the highest redshift as of now, also has $M_{\rm star}>M_{\rm Coma}$.
GS-z14-0 as well as GS-z14-1, another galaxy with $z>14$, seem to be on the SFMS at the redshift, and GS-z14-0 has much higher $M_{\rm star}$ than the simulated PCR galaxies.
GS-z14-1 is located at the projected distance of $6.2$~cMpc from GS-z14-0.
The redshift difference between these two galaxies is 0.42 which corresponds to $\sim60$~cMpc.
While this distance significantly exceeds the typical size of PCs at this redshift, \cite{Carniani:2024hb} suggests these galaxies could be part of an extended large-scale structure.
Deeper observations would shed light on the environment of GS-z14-0.

\subsection{UV magnitude and UV continuum slope\label{sec:muv}}

\begin{figure*}
\begin{center}
\includegraphics[width=\textwidth, bb=0 0 1400 1378]{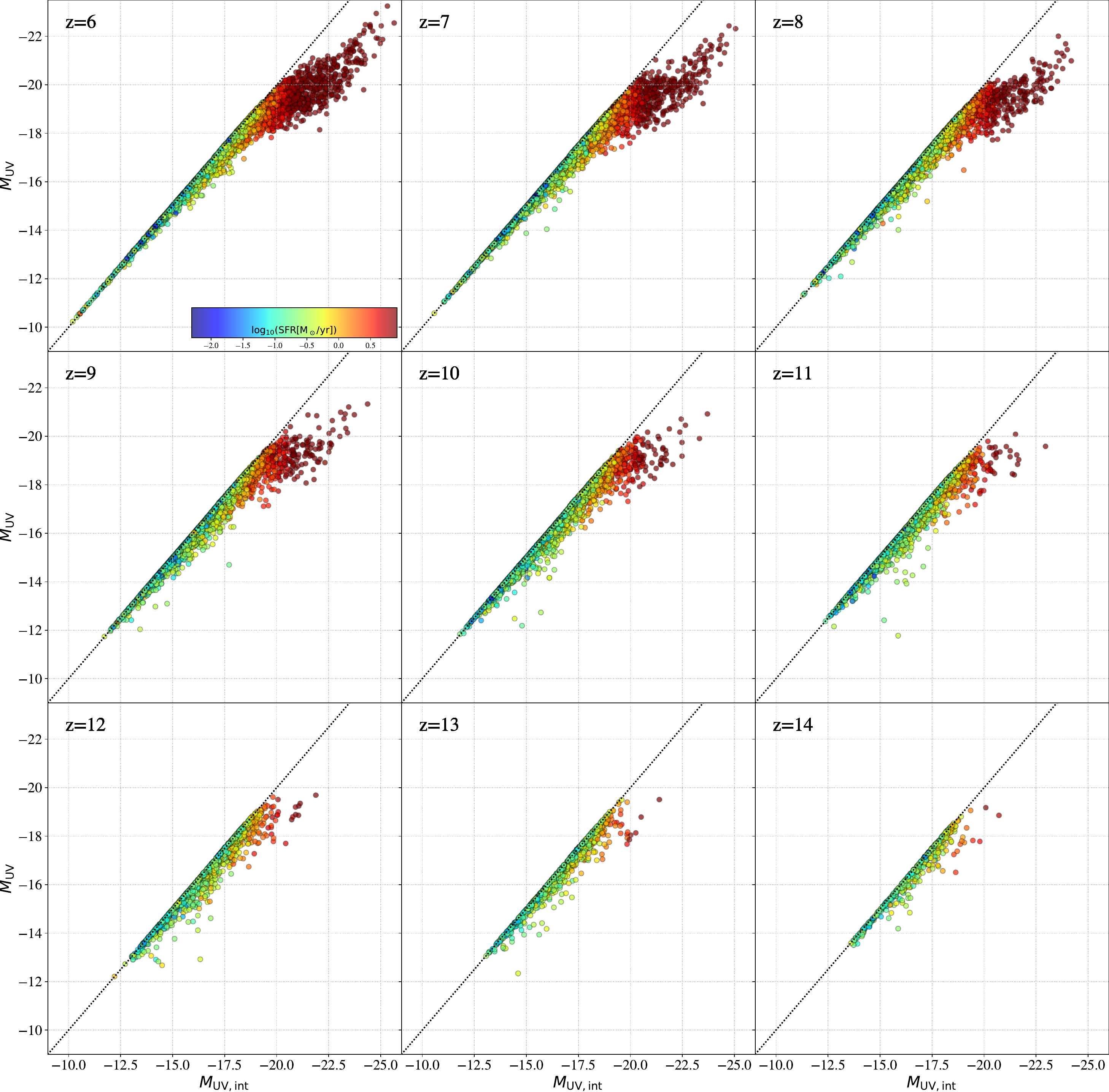}
\end{center}
\caption{
Relationship between the intrinsic UV magnitude and the observed UV magnitude of PCR galaxies estimated in the {\tt ART$^2$} calculation.
The color of each data point represents the SFR of the galaxy.
For intrinsically UV-bright galaxies with $M_{\rm UV, int}<-20$~mag, the observed UV magnitude can be fainter than the intrinsic value by more than 2 magnitudes due to dust attenuation.
Galaxies with higher SFRs exhibit stronger dust attenuation effects for a fixed $M_{\rm UV, int}$.
}
\label{fig:muv_art2}
\end{figure*}

\begin{figure*}
\begin{center}
\includegraphics[width=\textwidth, bb=0 0 1406 496]{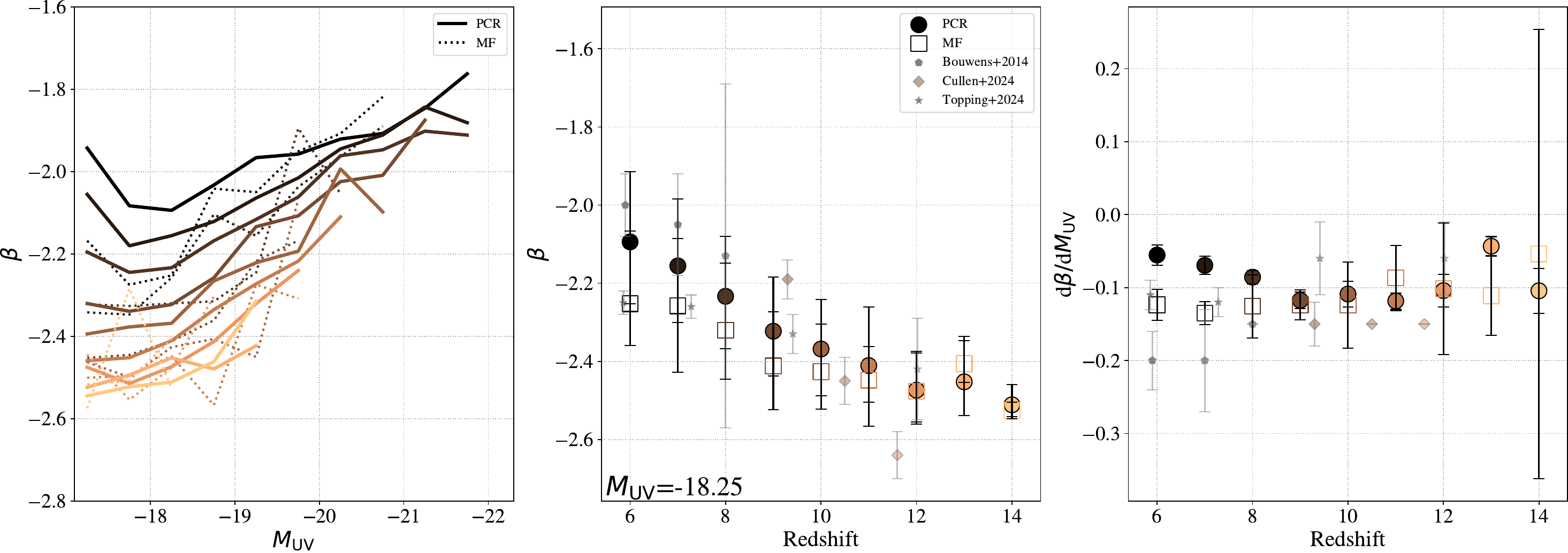}
\end{center}
\caption{
M$_{\rm UV}-\beta$ relationship and its redshift evolution.
Left: M$_{\rm UV}-\beta$ relationship for the PCR (solid line) and MF (dotted line) galaxies at $z=6-14$.
Center: Redshift evolution of $\beta$ of the PCR (circles), and MF galaxies (squares) with M$_{\rm UV}$ of $-18.5\sim-18$~mag.
Note that the $\beta$ observed is for objects with M$_{\rm UV}\sim -19$~mag.
Right: Redshift evolution of the slope of the M$_{\rm UV}-\beta$ relationship (d$\beta/$d$M_{\rm UV}$).
In the panels illustrating the redshift evolution of $\beta$ and d$\beta/$d$M_{\rm UV}$, filled circles and open squares represent the median values for PCR and MF galaxies, respectively.
The 25th and 75th percentiles are also shown for each population.
In the left panel, line colors, and in the center and right panels, symbol colors, are determined by redshift.
The center and right panels also show observed $\beta$ and d$\beta/$d$M_{\rm UV}$ \citep{Bouwens:2014cl,Cullen:2024lr,Topping:2024ao}.
}
\label{fig:muv_beta}
\end{figure*}
This subsection presents the properties of UV emission from simulated galaxies, which will be used in the subsequent analysis. A detailed investigation of dust properties in these simulated galaxies will be presented in future work.
Fig.~\ref{fig:muv_art2} compares intrinsic ($M_{\rm UV,int}$) and attenuated ($M_{\rm UV}$) UV magnitudes of the PCR galaxies.
Massive galaxies with $M_{\rm UV,int}<-20$~mag exhibit $\gtrsim 2$ magnitudes of stronger UV dust attenuation at redshifts $6-14$, with this attenuation decreasing at higher redshifts.
Consequently, when detecting galaxies with an observed UV magnitude of $M_{\rm UV}\sim-19$~mag, the sample may include not only galaxies with an intrinsic UV magnitude of $M_{\rm UV,int}\sim-19$~mag but also those with $M_{\rm UV,int}\sim-22.5$~mag.
Notably, for a given $M_{\rm UV}$, galaxies with higher SFRs tend to exhibit stronger UV attenuation.
This highlights the importance of radiative transfer calculations for accurate comparisons with observational data.

\subsubsection{$M_{\rm UV}-\beta$ relation}\label{sec:muv_beta}

The UV stellar continuum slope $\beta$ has been used to probe dust obscuration of star-forming galaxies.
$\beta$ is sensitive to dust attenuation as well as to the light-weighted age and metallicity of massive stars within galaxies.
However, at higher redshifts, dust attenuation is expected to play an essential role in determining $\beta$ due to the lack of time for significant chemical enrichment and variation in stellar populations \cite[e.g.,][]{Tacchella:2022ad}.
Extensive observations revealed a redshift evolution of $\beta$ where $\beta$ is $\sim-1.8$ at $z\sim2-4$, $\sim-2$ at $z\sim6$, and $\sim-2.5$ at $z\sim10$ for galaxies with $M_{\rm UV}\sim-19$~mag \citep[e.g.,][]{Bouwens:2009rp,Bouwens:2014cl,Finkelstein:2012oj,Topping:2024ao,Cullen:2024lr}.
This observed trend aligns with the current model of galactic evolution, suggesting that galaxies in the past were characterized by lower metallicities and dust depletion.

The left panel of Fig.~\ref{fig:muv_beta} shows the redshift evolution of the $M_{\rm UV}-\beta$ relation at $z=6-14$.
We measure $\beta$ by excluding the prominent dust absorption feature at $2175$~\AA, following the method of \cite{Calzetti:1994ab}.
Our simulations roughly reproduce the observed range of $\beta$ at $z\sim6-12$.
In this figure, the PCR and MF galaxies are represented by solid lines+circles and dotted lines+squares, respectively.
We can see that the UV bright galaxies tend to have a redder $\beta$ in the redshift range we explored.
The normalization of the $M_{\rm UV}-\beta$ relation evolves with time where the galaxies at higher redshifts have a bluer UV continuum for fixed $M_{\rm UV}$, as found in previous observations.
A notable feature of the $M_{\rm UV}-\beta$ relation is an upturn in $\beta$ at $M_{\rm UV} > -18$~mag.
This upturn can be attributed to several factors.
First, in low-mass galaxies, prolonged suppression of star formation due to SN feedback can lead to redder $\beta$, resulting from the presence of older stellar populations.
Second, the contribution of highly dust-obscured massive galaxies can also contribute to the observed increase in $\beta$ in this luminosity range.
Furthermore, in the less bright UV regime, PCR galaxies exhibit redder $\beta$ compared to MF galaxies at $z\lesssim 11$.
This suggests two possible explanations: (1) a higher fraction of highly dust-obscured massive galaxies within PCRs, and (2) accelerated dust formation in lower-mass galaxies residing within the PC environment.
\cite{Roberts-Borsani:2024ka} reported the same tendency where massive galaxies tend to have shallower $\beta$ than less-massive counterparts using the JWST/NIRSpec data.

The center panel of Fig.~\ref{fig:muv_beta} shows the redshift evolution of $\beta$ of galaxies with $-18.5~{\rm mag}\leq M_{\rm UV}<-18.0$~mag.
$\beta$ of both PCR and MF galaxies decreases with increasing redshift, suggesting that distant galaxies predominantly contain less dust.
Again, we can see that PCR galaxies exhibit redder $\beta$ than MF galaxies.
The redshift evolution of $\beta$ is consistent with observations \citep[e.g.,][]{Bouwens:2009rp,Bouwens:2014cl,Finkelstein:2012oj,Topping:2024ao,Cullen:2024lr}.
However, \cite{Saxena:2024id} recently reported a mild increase of $\beta$ at $z>9.5$, which is inconsistent with our results.
They investigated $\beta$ of 295 galaxies using spectroscopic data obtained with JWST NIRSpec/PRISM.
They found a mild decrease in $\beta$ with decreasing $M_{\rm UV}$ and increasing redshift at $5.5<z<9.5$ while an increase of $\beta$ at $z>9.5$.
They claimed that the redder $\beta$ of galaxies at $z>9.5$ requires rapid dust build-up in the very early Universe or a significant contribution from nebular continuum to UV continuum emission from galaxies.
Since the number of galaxies with measurements of $\beta$ is currently small ($<20$) at $z>9.5$, larger samples of galaxies at high redshift are needed to constrain the redshift evolution of $\beta$ robustly.

Some recent studies find an extremely blue $\beta<-2.7$ galaxies at $z\gtrsim7$ \citep[e.g.,][]{Topping:2022od,Topping:2024ao,Bouwens:2023vn,Nanayakkara:2023wx,Cullen:2024lr,Austin:2024ae,Yanagisawa:2024it}, potentially indicating a high fraction of Lyman-continuum leakage.
The intrinsic stellar spectra of young, metal-poor galaxies define the bluest possible UV slope, $\beta\sim-3.0$ \citep[e.g.,][]{Bouwens:2010xp}.
However, nebular continuum emission from ionized gas reddens this slope to $\beta\sim-2.6$ \citep[e.g.,][]{Raiter:2010zc,Stanway:2016ss,Cameron:2023ou,Katz:2024vm,Saxena:2024id}.
In our {\tt ART$^{2}$} calculation, there is no such an extremely blue galaxy since nebular emission is taken into account.

The right panel of Fig.~\ref{fig:muv_beta} shows the redshift evolution of the slope of the $M_{\rm UV}-\beta$ relation, i.e., d$\beta/$d$M_{\rm UV}$.
d$\beta/$d$M_{\rm UV}$ is almost constant with a value of $\sim-0.1$ over the redshift range that we explored.
This is consistent with recent JWST observations showing a weak redshift dependence of d$\beta/$d$M_{\rm UV}\sim-0.1$ - $-0.2$ \citep{Cullen:2024lr,Austin:2024ae}.
Note that a significant scatter also exists in observed d$\beta/$d$M_{\rm UV}$ \citep{Bouwens:2012en,Bouwens:2014cl,Topping:2024ao,Roberts-Borsani:2024ka}.
\cite{Austin:2024ae} reported a flat d$\beta/$d$M_{\rm UV}$ value of $\sim0.03$ at $z\sim7$ and suggested a new population of low-mass, faint galaxies reddened by dust produced in the stellar wind of asymptotic giant branch stars or carbon-rich Wolf-Rayet stars.
The observed scatter in d$\beta$/$dM_{\rm UV}$ calls for further investigation to understand the underlying physical mechanisms driving the variations in dust attenuation properties across the galaxy population.

\subsection{Galaxy overdensity}\label{sec:overdensity}

We compare the galaxy overdensity measured with simulation data to those reported in observed PC candidates.
Galaxy overdensity is calculated as
\begin{equation}
    \delta=\frac{n-\bar{n}}{\bar{n}},
	\label{eq:od}
\end{equation}
where $n$ represents the number of galaxies within a specified volume and $\bar{n}$ represents $n$ of the normal field.

We investigate the redshift evolution of $\delta$ in PCs of Coma-type clusters, exploring its dependence on the search volume and the data sensitivity.
We considered seven cases of the limiting absolute ultraviolet (UV) magnitude at $1500$~\AA, $M_{\rm UV}$ of $-21, -20, -19, -18, -17, -16, -10$~mag.
Crucially, this analysis utilizes the dust-attenuated $M_{\rm UV}$ values.
The UV luminosity function at the EoR has been extensively studied using both ground-based and space telescopes \citep[e.g.,][]{Ellis:2013gd,Bowler:2015xc,Finkelstein:2015um,Bouwens:2021kl,Bouwens:2022qm,McLure:2013hc,McLeod:2016ts}.
The faint-end $M_{\rm UV}$ in these surveys is typically $\sim-16$~mag at $z\sim6-7$ \citep[e.g.,][]{Bouwens:2015ol} and $\sim-17$~mag at $z\sim8-10$ \citep[e.g.,][]{McLeod:2016ts,Bouwens:2021kl}.
With the power of the gravitational-lensing effect, these values increase to $\sim-13$~mag at $z\sim6-8$ and $-16$~mag at $z\sim9$ \citep{Bouwens:2022qm}.

We employ cylindrical search volumes with a fixed height of 20~cMpc~$h^{-1}$ and varying radii from 1 to 10~cMpc~$h^{-1}$, resulting in volumes ranging from 63 to 6283~cMpc$^3$~$h^{-3}$.
To calculate $\delta$, we project the 3D galaxy distribution onto the XY, XZ, and YZ planes, deriving three $\delta$ values for each galaxy within a PC snapshot.

The average number of galaxies in normal fields ($\bar{n}$) is estimated by using the MF data in the FOREVER22.
Three distinct MFs were considered in the FOREVER22 project.
For each MF, the total number of galaxies brighter than the limiting UV magnitude was counted within its entire volume.
Finally, the 3D number density of galaxies in each MF was calculated for each redshift.

\subsubsection{Overdensity dependence on the search volume and limiting magnitude of the data and its evolution}

\begin{figure*}
\begin{center}
\includegraphics[width=\textwidth, bb=0 0 1396 938]{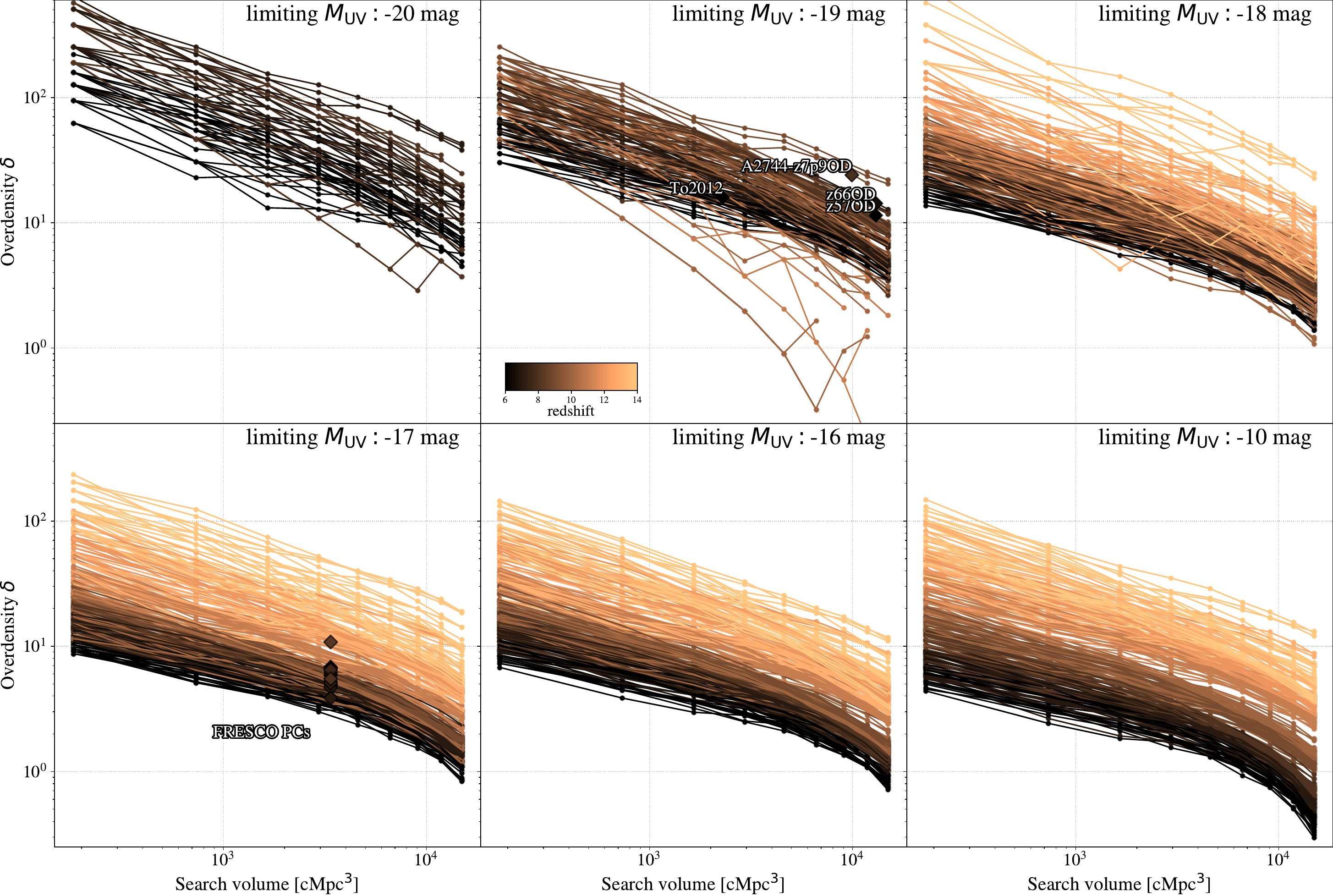}
\end{center}
\caption{$\delta$ as a function of search volume and limiting $M_{\rm UV}$. Filled diamonds represent observational data from  \protect\cite{Toshikawa:2012tm,Harikane:2019ss,Morishita:2023jh,Helton:2024fx}. For the panel for a limiting $M_{\rm UV}=-19$~mag, the labels indicate the names of the observed PCs: To2012 \protect\citep{Toshikawa:2012tm}; z66OD and z57OD \protect\citep{Harikane:2019ss}; A2744-z7p9OD \protect\citep{Morishita:2023jh}. In the panel for a limiting $M_{\rm UV}=-17$~mag, PCs from \protect\cite{Helton:2024fx} are shown without individual labels. Line and symbol colors are assigned according to redshift.}
\label{fig:od_art2}
\end{figure*}

Fig.~\ref{fig:od_art2} shows $\delta$ as a function of the search volume for each redshift and the limiting $M_{\rm UV}$.
Diamond symbols indicate observation with color-coded according to their redshifts \citep{Toshikawa:2012tm,Toshikawa:2014zm,Harikane:2019ss,Helton:2024fx}.
Our $\delta$ fairly matches well with observations if both the search volume and limiting magnitude of the data are (roughly) matched.
The exceptionally high $\delta$ of $\sim130$ was originally reported in the core of A2744-z7p9OD based on the photometric data \citep{Ishigaki:2016px}.
\cite{Morishita:2023jh} updated this value to $\delta\sim24$ using the spectroscopic data, which is shown in Fig.~\ref{fig:od_art2}.
A2744-z7p9OD lies slightly above our prediction.

$\delta$ increases with increasing redshift, decreasing search volume, and sensitivity of the data, i.e., lower limiting $M_{\rm UV}$.
These tendencies are consistent with previous studies.
This redshift dependence aligns with findings from \cite{Yajima:2022jl} that PCR galaxies exhibit higher stellar mass function normalization at higher redshifts, indicating accelerated galaxy formation and evolution compared to MF galaxies.
The correlation between $\delta$ and the limiting $M_{\rm UV}$ reflects the tendency of brighter galaxies to inhabit denser regions.
Fig.~\ref{fig:3d} demonstrates that massive galaxies tend to reside in denser regions than their less-massive counterparts.
This tendency is also seen in the observational study, where \cite{Harikane:2016hw} demonstrated a shorter correlation length for brighter galaxies using the Subaru/Hyper Suprime-Cam survey data.

\subsubsection{Overdensity dependence on $M_{\rm UV}$ and $\beta$ slope}\label{sec:delta_muv_beta}

\begin{figure*}
\begin{center}
\includegraphics[width=\textwidth, bb=0 0 1403 1376]{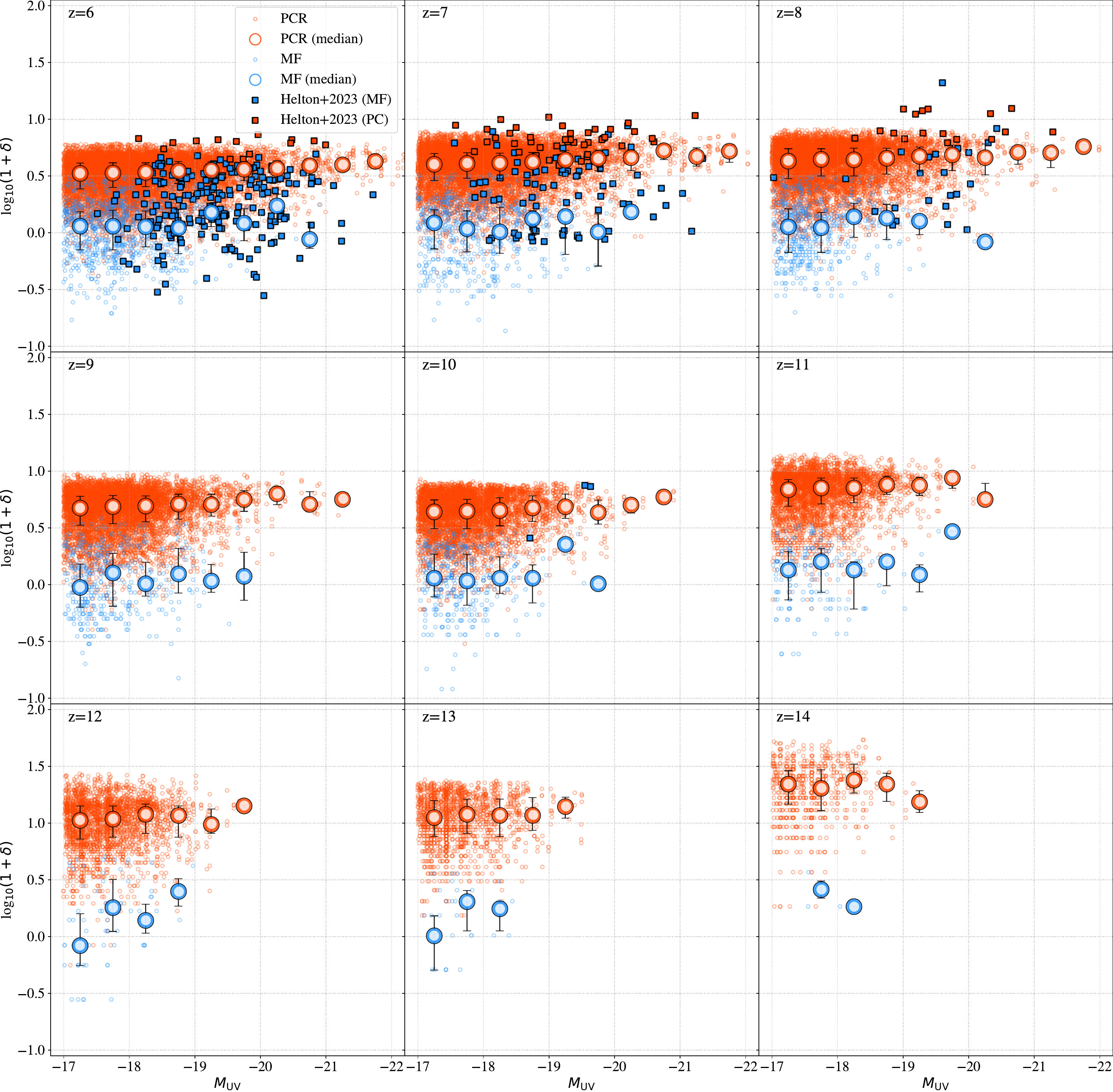}
\end{center}
\caption{
Overdensity ($\delta$) as a function of $M_{\rm UV}$ of galaxies in the PCR and the mean field (MF) at redshifts of $z=6-14$.
Small open circles represent individual FOREVER22 galaxies in the PCR (red) and the MF (blue).
Large open red and blue circles denote the median value of the PCR and MF galaxies, respectively.
The 25th and 75th percentiles for each population are also shown.
Red- and blue-filled squares outlined in black represent observational data from \protect\cite{Helton:2024fx}.
}
\label{fig:od_muv}
\end{figure*}

\begin{figure*}
\begin{center}
\includegraphics[width=\textwidth, bb=0 0 1403 1376]{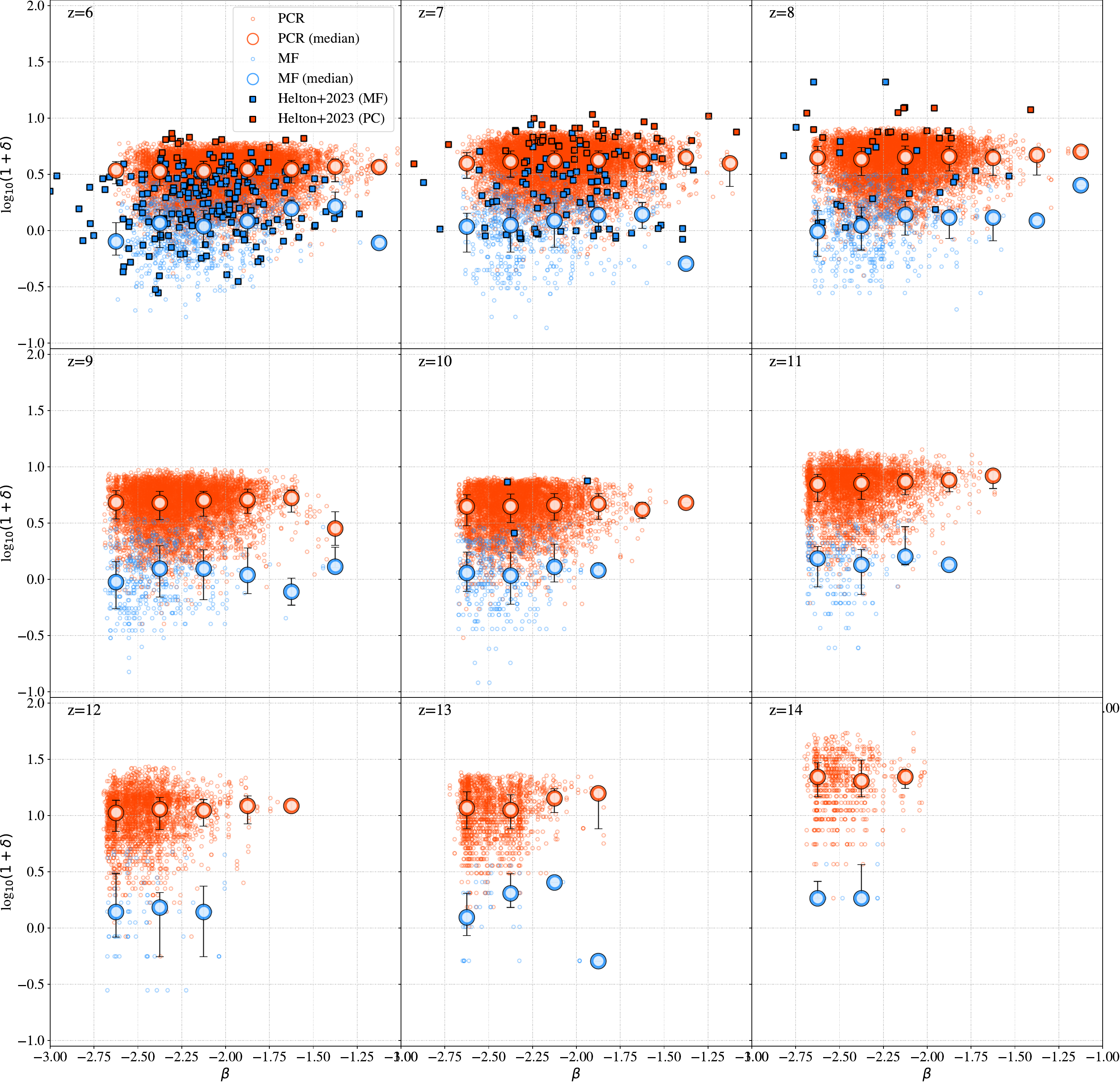}
\end{center}
\caption{
Overdensity ($\delta$) as a function of UV dust attenuation slope ($\beta$) of galaxies in the PCR and the MF at redshifts of $z=6-14$.
Symbol conventions are as in Figure~\ref{fig:od_muv}.}
\label{fig:od_beta}
\end{figure*}

\cite{Helton:2024fx} spectroscopically confirmed 17 PCs of H$\alpha$ and \oiii~emitters at $4.9<z<8.9$ in the GOOD-N and GOOD-S fields using JWST/NIRCam imaging data from JADES \citep{Eisenstein:2023ao} and JEMS \citep{Williams:2023ee} in addition to JWST/NIRCam wide field slitless spectroscopic data from FRESCO \citep{Oesch:2023qd}.
To identify PCs, they applied a friends-of-friends (FoF) algorithm to their 3D data, setting linking thresholds at $d_{\rm link}=500$~kpc for spatial proximity and $\sigma_{\rm link}=500~$km~s$^{-1}$ for redshift proximity.
They reported a tendency where the galaxies with a higher $\delta$ are bright at UV wavelength and have a redder UV slope, $\beta$.
For the comparison with \cite{Helton:2024fx}, we use $\delta$ which is calculated for the case of $M_{\rm UV, lim}=-17$~mag, and the search volume of 3,079~[cMpc$^3$/$h^3$].

We compare the $\delta-M_{\rm UV}$ relation from our simulations to observational data presented in \cite{Helton:2024fx}.
In Fig.~\ref{fig:od_muv}, we shows the PCR (red symbols) and the MF (blue symbols) galaxies separately.
The simulated and observed galaxies are denoted with open cycles and with filled squares, respectively.
The large circles are the median values of $\delta$ for simulated galaxies with similar $M_{\rm UV}$.
Our simulations generally reproduce the observed trend of higher $\delta$ values for the PCR galaxies than the MF galaxies (as per the definition), although with some overlap.
Notably, our simulations accurately capture the absence of UV-bright galaxies with low $\delta$ values, consistent with observational findings.
This confirms the observed trend of UV-bright galaxies occupying denser environments.

Fig.~\ref{fig:od_beta} shows the dependence of $\delta$ on $\beta$.
There is no significant dependence of $\delta$ on $\beta$ if considering PCR and MF galaxies separately.
A trend emerges where galaxies within denser regions tend to possess redder average UV slopes as seen in \cite{Helton:2024fx} when considering both PCR and MF populations equally.
This trend is attributed to the fact that there are more massive galaxies in PCRs than MFs (Fig.~\ref{fig:mstarsfr_mf}) and massive galaxies exhibit consistently redder UV slopes than less-massive galaxies (Fig.~\ref{fig:muv_art2}).

\section{What are the reliable methods for identifying a Coma-type protocluster at the Epoch of Reionization?}\label{sec:dis_pc}

$\delta$ has been employed to identify and characterize PCs.
However, relying solely on $\delta$ may not effectively distinguish true PCs from contaminants.
We find that some MF galaxies exhibit a high $\delta$ values, resembling those of PCR galaxies, while PCR galaxies generally exhibit higher $\delta$ values than those in the mean field.
Furthermore, $\delta$ is an evolving quantity that depends on both search volume and observational sensitivity.
Thus, direct comparisons of $\delta$ values between PC candidates from different studies can be challenging when considering data sets with varying sensitivities and PC selection criteria, as these factors can significantly influence the measured overdensity.

\cite{Helton:2024fx} reported a higher number density of Coma-type PCs, $n_{\rm Coma} = 2.2 \times 10^{-5}$ cMpc$^{-3}$, than the predicted values for Coma-type and Fornax-type clusters, $1.8 \times 10^{-7}$ cMpc$^{-3}$ and $6.1 \times 10^{-6}$ cMpc$^{-3}$, respectively, based on semi-analytic galaxy formation models from \cite{Chiang:2013eb}.
Furthermore, recent large-scale cosmological galaxy simulations identified approximately 100 clusters with $M_{\rm halo}>10^{15.0}$~M$_\odot$ within a simulation volume of $1003.8^3$~Mpc$^3$ \citep{Nelson:2024ef}, resulting in $n_{\rm Coma}\sim 10^{-7}$~cMpc$^{-3}$.

Another key difference of galaxies in PCR and MF lies in the mass of the most massive galaxy ($M_{\rm star,max}$) within the system.
Enhanced galaxy interactions and gas accretion within the denser environments of PCs likely facilitate the formation of more massive galaxies.
While the $M_{\rm star}$-SFR relation does not exhibit significant differences between PC and MF galaxies, we observe that PC galaxies extend to higher $M_{\rm star}$ and SFR regimes. 
Due to the significant scatter in the mass assembly histories of dark matter halos, massive high-redshift systems do not necessarily evolve into the most massive systems in the local universe \citep{Remus:2023mw,Lim:2024um}.
$M_{\rm star, max}$ offers an indicative, rather than definitive, prediction of the system's evolutionary trajectory.
Therefore, the presence of massive galaxies with $M_{\rm star} > M_{\rm Coma}$ or $M_{\rm star} > M_{\rm Fornax}$, in conjunction with elevated overdensity ($\delta$), further strengthens the identification of PCs as potential progenitors of Coma-type and Fornax-type galaxy clusters, respectively.

We compare the observed $M_{\rm star, max}$ of Helton's PCs (one PC at $z\sim6$, nine PCs at $z\sim7$, and three PCs at $z\sim8$) to $M_{\rm Coma}$ and $M_{\rm Fornax}$ at each redshift.
Considering error margins, we find that two out of the 13 PCs at $z\sim8$ have a member galaxy with $M_{\rm star,max}>M_{\rm Coma}$ and eight out of the 13 PCs have $M_{\rm star,max}>M_{\rm Fornax}$.
Therefore, the observed value of $n_{\rm Coma}$ in \cite{Helton:2024fx} can be overestimated by nearly an order of magnitude. However, even after accounting for this potential overestimation, the observed value remains significantly higher than the theoretical prediction.
We should continue to investigate the causes of this discrepancy between observations and theory.

\begin{figure}
\begin{center}
\includegraphics[width=0.5\textwidth, bb=0 0 501 488]{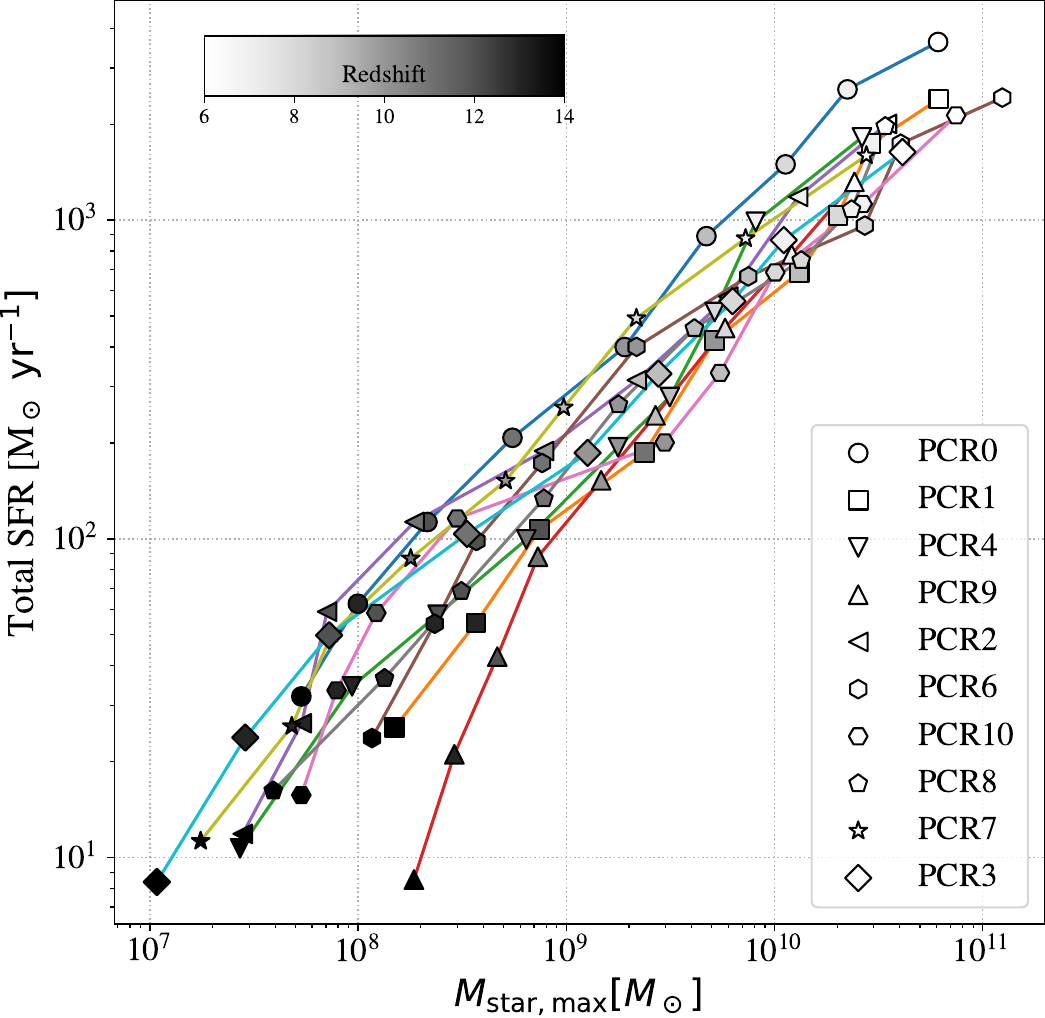}
\end{center}
\caption{
Relationship between the stellar mass of the most massive galaxy and the total SFR within a central sphere with a radius of 10~cMpc in a PCR calculation box.
}
\label{fig:mstarmax_totalsfr}
\end{figure}

Furthermore, measuring $M_{\rm star,max}$ within a PC can serve as a proxy for estimating the total SFR within the entire overdense region.
Fig.~\ref{fig:mstarmax_totalsfr} illustrates the correlation between $M_{\rm star,max}$ and the total SFR integrated within a 10 cMpc radius (SFR$_{\rm total}$).
Although a significant scatter exists in the $M_{\rm star,max}$-SFR$_{\rm total}$ relation, especially at redshifts greater than 10, this relationship provides valuable guidance for optimizing follow-up observations of PC candidates, not only in the rest-frame UV-optical but also in sub-millimeter wavelengths, to effectively detect member galaxies. 
Substantial obscured star formation has been reported in some PCs, even at the EoR, as exemplified by SPT0311-58, a massive PC hosting two heavily dust-obscured galaxies with extreme SFRs \citep{Strandet:2017gm}.
In the Appendix section, Fig.~\ref{fig:mstarmax_totalsfr_eachz} presents the $M_{\rm star,max}$-SFR$_{\rm total}$ relation for individual PCRs at different redshifts.

\section{Summary}\label{sec:summary}

We investigate the galaxy overdensity and star-formation properties of galaxies in the progenitor of Coma-type galaxy clusters at redshifts from 6 to 14 using the data of cosmological hydrodynamic simulation, FOREVER22 \citep{Yajima:2022jl}.
To compare with observations, we perform radiative transfer calculations on simulated data using {\tt ART$^2$} \citep{Yajima:2012ty}, incorporating dust extinction.

The primary findings are:
\begin{description}
 \item[{\bf $M_{\rm star}-$SFR relationship} (section~\ref{sec:overdensity})]
 Simulated galaxies in both PCs and mean fields occupy the same star formation main sequence at each redshift (Fig.~\ref{fig:mstarsfr_mf}). However, PCR galaxies span a wider range of stellar masses and SFRs than MF galaxies, with some PCR galaxies exhibiting higher values than any of the MF galaxies. The stellar masses of massive PCR galaxies are found to be greater than the progenitor masses of Coma-type galaxy clusters when estimated using the extended Press-Schechter formalism. Our simulations accurately reproduce the observed $M_{\rm star}-$SFR relation between redshifts six and 14, albeit with a discrepancy for a subset of observed galaxies characterized by lower star formation rates for fixed stellar masses (Fig.~\ref{fig:mstarsfr}).

 \item[{\bf Galaxy overdensity} (section~\ref{sec:overdensity})]
 We find that galaxy overdensity ($\delta$) is sensitive to both search volume and limiting magnitude, with smaller regions and shallower observations yielding higher values (Fig.~\ref{fig:od_art2}). $\delta$ increases with redshift, suggesting higher density contrast in the early Universe. These trends imply that PC cores are exceptionally dense, massive galaxies exhibit stronger clustering, and galaxy formation progresses from dense to less dense environments.
 When comparing overdensity values across different studies, it is essential to account for variations in search volume and data sensitivity.
 
  \item[{\bf UV magnitude and UV continuum slope} (sections~\ref{sec:muv}; \ref{sec:delta_muv_beta})]
 Massive galaxies exhibit stronger UV attenuation, with a maximum difference of over two magnitudes (Fig.~\ref{fig:muv_art2}). The UV continuum slope ($\beta$) of galaxies becomes redder over time (Figs~\ref{fig:muv_beta} left and center). PCR galaxies display redder $\beta$ values compared to field galaxies, primarily attributed to their higher fraction of massive members. Our simulations do not show a significant redshift evolution in the slope of $M_{\rm UV}-\beta$ relationship (Fig.~\ref{fig:muv_beta} right). Our simulations fairly reproduce the observed correlation between $\delta$, $M_{\rm UV}$, and $\beta$, where higher $\delta$ values correspond to brighter galaxies in UV with redder $\beta$ slopes (Fig.s~\ref{fig:od_muv} and \ref{fig:od_beta}).

\end{description}

Considering the findings presented above, we recommend identifying PCs based on not only $\delta$ but also the mass of their most massive member. Determining the stellar mass of the most massive galaxies within these regions allows for estimating the total SFR of the overdensity, which is useful for future observations targeting dark member galaxies.

\section*{Acknowledgements}
We thank the anonymous referee for their insightful comments and suggestions that significantly improved the manuscript.
KMM gratefully acknowledges Dr.~J.~Helton and Dr.~F.~Sun for supplying their observational data.
The numerical simulations were performed on the computer cluster, XC50 in NAOJ, and Trinity at the Center for Computational Sciences in University of Tsukuba. This work is supported in part by MEXT/JSPS KAKENHI Grant Numbers 17H04827, 20H04724, and 21H04489 and JST FOREST Program, Grant Number JP-MJFR202Z (HY).


\section*{Data Availability}

The data in this paper will be shared on reasonable request to the corresponding author. 



\bibliographystyle{mnras}
\bibliography{myref_f22} 




\appendix

\section{Some extra material}

\begin{figure*}
\begin{center}
\includegraphics[width=\textwidth, bb=0 0 1386 1376]{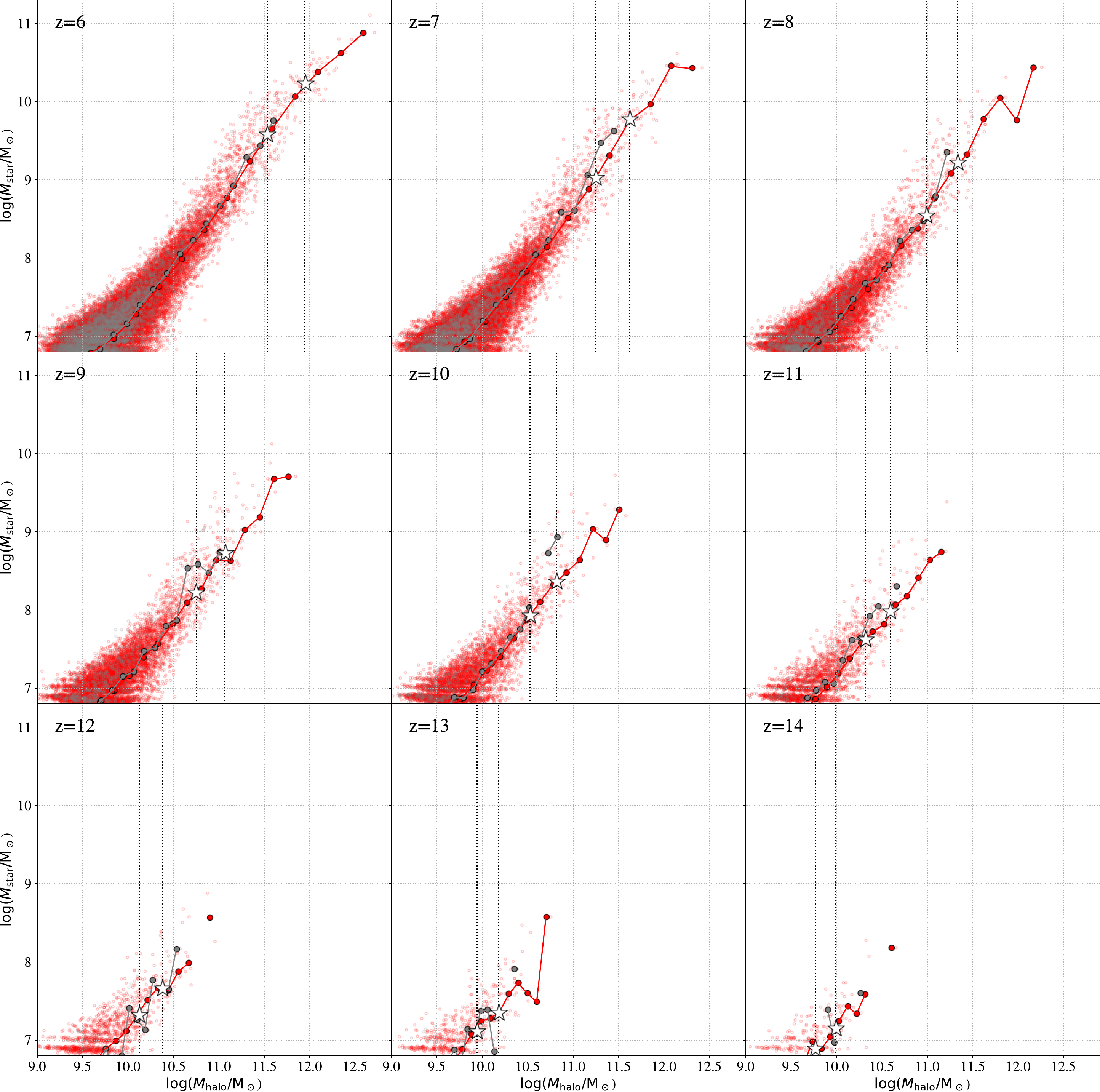}
\end{center}
\caption{
$M_{\rm halo}-M_{\rm star}$ relations of FOREVER galaxies from redshift six to 14.
Individual PCR and MF galaxies are indicated with open red and gray circles, respectively.
The filled red and gray circles indicate the medians of $M_{\rm star}$ for the PCR galaxies with similar $M_{\rm halo}$.
Dotted and dot-dashed lines indicate the progenitor masses of Fornax-type ($M_{\rm Fornax}$) and Coma-type galaxy clusters ($M_{\rm Coma}$) at each redshift, respectively.
White stars indicate the median $M_{\rm star}$ value for galaxies with $M_{\rm halo}=M_{\rm Fornax, h}, M_{\rm Coma, h}$, which are shown in Figs~\ref{fig:mstarsfr} and \ref{fig:mstarsfr_mf}.
The progenitor masses for the Fornax-type and Coma-type clusters are calculated according to the extended Press-Schechter formalism by assuming that the masses of Fornax-type and Coma-type clusters are $10^{14}$~M$_\odot$ and $10^{15}$~M$_\odot$, respectively.
}
\label{fig:mhalomstar}
\end{figure*}

\begin{figure*}
\begin{center}
\includegraphics[width=\textwidth, bb=0 0 1408 1379]{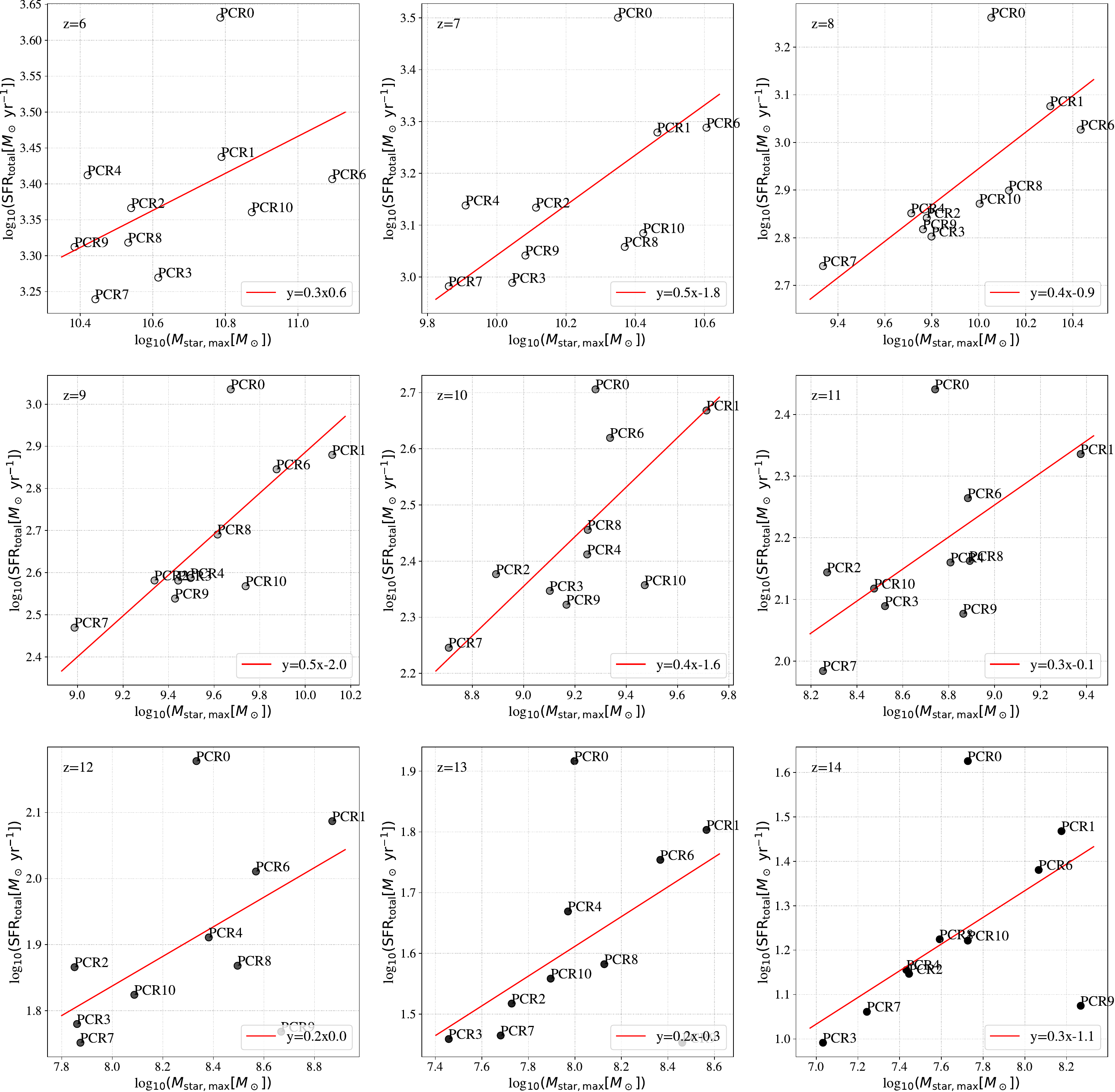}
\end{center}
\caption{
Relationship between the stellar mass of the most massive galaxy and the total SFR within a central sphere with a radius of 10 [cMpc] in the PCR calculation box at each redshift.
}
\label{fig:mstarmax_totalsfr_eachz}
\end{figure*}

In this section, we present additional figures.
Fig.~\ref{fig:mhalomstar} shows the $M_{\rm halo}-M_{\rm star}$ relations of the PCR (red) and MF galaxies (gray) at $z=6-14$.
These relations are used to estimate the $M_{\rm star}$ for the progenitor halo of Fornax-type and Coma-type galaxy clusters at each redshift.
We adopted the median $M_{\rm star}$ for halos with similar $M_{\rm halo}$ as the progenitor stellar masses.

Fig.~\ref{fig:mstarmax_totalsfr_eachz} shows the $M_{\rm star, max}-$SFR$_{\rm total}$ relation of ten PCRs at $z=6-14$.
SFR$_{\rm total}$ is calculated as the sum of SFR within a central sphere with a radius of 10~cMpc in the PCR calculation box.



\bsp	
\label{lastpage}
\end{document}